# WHAT CAN SOCIAL MEDIA TEACH US ABOUT PROTESTS? ANALYZING THE CHILEAN 2011-12 STUDENT MOVEMENT'S NETWORK EVOLUTION THROUGH TWITTER DATA

García, C., Chauveau, P., Ledezma, J. & Pinto, Ma.

Pontificia Universidad Catolica de Chile's School of Business
4860 Vicuña Mackenna Avenue
Santiago, Metro Region, 7820436, Chile
e-mail: cgarciah@uc.cl

**ABSTRACT**

Using social media data—specially twitter—of the Chilean 2011-12 student movement, we study their social network evolution over time to analyze how leaders and participants self-organize and spread information. Based on a few key events of the student movement's timeline, we visualize the student network trajectory and analyze their structural and semantic properties. Therefore, in this paper we: i) describe the basic network topology of the 2011-12 Chilean massive student movement; ii) explore how the 180 key central nodes of the movement are connected, self-organize and spread information. We contend that this social media enabled massive movement is yet another manifestation of the network era, which leverages agents' socio-technical networks, and thus accelerates how agents coordinate, mobilize resources and enact collective intelligence.

## I. CONTEXT: THE NETWORKED SOCIETY & STUDENT MOVEMENT

The world has been in a process of structural transformation for over two decades. This process is multidimensional, associated with the emergence of a new technological paradigm, based on information and communication technologies, which emerged in the 1970s and are spreading around the world. Society shapes technology according to the needs, values, and interests of people who use the technology: "Technology does not determine society: it expresses it. But society does not determine technological innovation: it uses it" (Castells, 1996/2000, p.15). Technology appears as a key tool in social processes, "it is people´s usage of technology – not technology itself – that can change social process" (Earl & Kimport, 2011, p.14). Furthermore, information and communication technologies are particularly sensitive to the effects of social uses on technology. Internet has evolved "from organizational business tool and communication medium to a lever of social transformation as well" (Castells, 2002, p.143). However, technology is a necessary, albeit not sufficient condition for the emergence of a new form of social organization based on networking, that is, the diffusion of networking in all realms of activity on the basis of digital communication networks.

Changes in the technologies, economic organization, and social practices of production in this networked environment have shaped new opportunities for how we create and exchange information, knowledge, and culture. These changes "have increased the role of nonmarket and nonproprietary production, both by individuals alone and by cooperative efforts in a wide range of loosely or tightly woven collaborations'' (Benkler, 2005, p.5). These changes have increased the thresholds of available freedom for users, consumers and citizens, and thus hold great practical promise: as a dimension of individual freedom; as a platform for better democratic participation; and as a medium to foster a more critical and self-reflective culture. In this sense, the Chilean networked student movement is an opportunity to study the use of social media—specially Twitter—as a mean to enact collective intelligence and thus, to critique and challenge the power of an incumbent government.

We describe the Chilean networked student movement as it has been unfolding during the selected period of 2011-12. We explore the network topology, emerging leaders, influencers and connectors by analyzing diachronic new media use and content in Twitter. In addition, we tracking the events of the student movement and built a timeline of relevant occurrences. The aim of the paper is to fit the structure of the network with the timeline of events, to understand its dynamics. We contend the current networked smart movement in Chile is yet another example of collective intelligence enabled by social media in turbulent 2011.

On the one hand, we built a timeline of the student movement, to analyze its evolution and network dynamics by tracking the selected events of the student movement between June 2011 and October 2012. On the other hand, we queried a global Twitter database by using Condor, social network/media analysis software to identify the central/peripheral/influential nodes and related metrics of the Chilean student movement during the selected period of the timeline.

The network society, in the simplest terms, is a social structure based on networks operated by information and communication technologies based in microelectronics and digital computer networks that generate, process, and distribute information on the basis of the knowledge accumulated in the nodes of the networks. A network is a formal structure (Monge & Contractor, 2003). A social network is a finite set of actors and the relationship between them, represented by a system of interconnected nodes (Wasserman & Faust, 1994). Following the Internet design, societies, organizations and movements have evolved from centralized to decentralized and distributed networks. This decentralization and democratization of decision-making has impacted businesses, governments and society at large (Malone, 2004). As the network society diffuses, and new communications technologies expand their networks, there is an explosion of horizontal networks of communication, quite independent from media business and governments, that allows the emergence of self-directed mass communications (Castells, 2005), Smart Mobs (Rheingold, 2002), participatory cultures (Jenkins 2006) and collaborative innovation networks (Gloor, 2006). It is mass communication because it is diffused throughout the Internet, so it potentially reaches whole countries and regions and eventually the whole planet. It is self-directed because it is often initiated by individuals or groups, bypassing the official media system. The explosion of user-generated content, often referred to as web 2.0, includes blogs, wikis, videoblogs, podcasts, social networking sites, streaming, and other forms of interactive, computer to computer communication and sets up a new system of global, horizontal communication networks. "The dominant institutions of society no longer have the monopoly of mass-communication networks" (Castells, Fernández-Ardèvol et al., 2007, p.213), for the first time in history, people can communicate with each other without going through the channels set up by the institutions of society for socialized communication.

Several studies have focus on the relationship between social media and social and political movements. Traditional social movement theory, albeit offer a valuable insights to have a comprehensive understanding of the actual social movements around the world, media activism have some shortcomings in existing conceptualizations (Carroll, 2006). The online diffusion of information have been the focus to understanding online protest and other expression of social movement, but it is not the only one (Earl & Barbara, 2010). Mass self-communication contributes to a new public sphere where social actors can exercise autonomy and promote change in ways that were previously impossible. (Al-Ani, Mark, Chung, & Jones, 2012).

What is the key that pushed the whole country to get together and demand a better education, what actually kept the movement awake, are questions to be addressed. Recent research had put their attention in the specific relationships between Internet tools and its role as an enhancing social request. Blogs played a crucial role within the context of the Egyptian revolution of early 2011 (Al-ani et al., 2012). By Reporting events and commenting on the Blogs, bloggers create a "counter-narrative" against the incumbent demands. These narratives were displayed to all country and at international level, as a free platform to manifest ideology and freedom of speech, without a fear of violence and intimidation.

How blogs are written during a conflict, represent a collective intelligence of the event, and the content of Blogs, can be viewed as an indicator of the state of population (Mark et al., 2012). In this sense, it seems that social media not only play a diffusion role, but is a space where reality is socially constructed (Berger & Luckmann, 1966). In particular, Internet appears as a communication medium, but also as the "infrastructure of a given organizational form: the network" (Castells, 2002, p. 139).

These new socio-technical conditions present both opportunities and challenges to the 'organizing process' (Weick, 1995; Malone, 2004) as well as to the democracy and society at large. Our research was motivated to understand the student revolution from the perspective of social media, to enhancing the comprehension of social phenomena in the networking era.

There are certain characteristics of Internet that make possible the unfolding of online mobilizations (Castells, 2002; Earl and Kimport, 2011). On the one hand, it allows the use of communication systems to transmit ideas and cultural values that sustain networked social movements. On the other hand, Internet is a tool that helps to coordinate and to organize a great number of people through a cheap, efficient and effective way. Thus, it allows to generate online protests that are articulated in the web, without previous meetings.

Earl and Kimport (2011) identify "three instances that represents a continuum of the web protest": e-mobilizations, e-tactics and e-movements.

> In e-mobilizations, online tools are used to facilitate the sharing of information in the service of an offline protest action, to bring people into the streets for face to face protest. In e.movements, organization of and participation in the movement occurs entirely online. Finally, e-tactics may include both off and online components (p. 12).

The Chilean student movement can be classified as an e-mobilization, because the main events happen in the streets, when people get together to demonstrate.

## II. THE CHILEAN EXPERIMENT: FROM THE 2006 PENGUIN REVOLUTION TO THE 2011-12 STUDENT NETWORKED AND GLOBAL MOVEMENT

The "Penguin Revolution" was the first student citizen movement in Chile that was completely independent from existing political parties. It originated during the first months of Michelle Bachelet Administration in May of 2006. Furthermore, the protagonists of this revolution were students, who were mainly from public high schools and between 15 and 18 years old. Street protests and national strikes conducted by them caused a big impact in the public opinion, mainly because of the appearance of students as relevant political actors, which was not seen since the dictatorship of Augusto Pinochet in the '80s. The student's main demand was to improve the quality of public education and to assure a fair educational system. Finally, during June of 2006, President Bachelet, who has just got in power with a citizen-oriented government mantra, proceeded to attend the short term demands of the student movement, in a backdrop of national strikes and deep critiques to the government. Moreover, not only this student network organization accomplished the way out of the education minister, but also achieved the modification of the Organic Constitutional Law of Education (LOCE). One of the key strengths of this early student movement was the civic use of internet-based tools, that is, both old and new media platforms by the different constituencies embracing this educational-oriented smart mob.

The new 2011-12 upsurge of popular unrest comes 18 months after a center-right president, Sebastián Piñera, took office. Before him, the center-left Concertación had ruled for a relatively tranquil 20 years, overseeing a long and delicate transformation to democracy with a few exceptions, such as the 2006 Penguin Revolution mentioned above. In 2011, thousands of high school and university students (some of them being 'grown-up penguins') marched through the streets of the capital and other major cities demanding a radical overhaul of the education system.

The 2011 student marches were far larger than those organized by other protest groups. On several occasions, the marches have drawn 100,000 people on to the streets. At the heart of the students' anger is the perception that Chile's education system is grossly unfair -- it gives rich students access to some of the best schooling in Latin America while dumping poor pupils in shabby, under-funded state schools. Out of the 65 countries that participated in the PISA tests, Chile ranked 64th in terms of segregation across social classes in its schools and colleges. Another relevant topic is the financing of the education. Chile is the OECD country with the highest share or private expenditure on educational institutions, with a 41%, versus the 16% OECD average (OECD, 2012). "The fact that Chilean students, or their families, have to directly pay a substantial share of the costs of their courses has focused attention on the value for money that they receive and their chances of finding a worthwhile career after graduation" (OECD, 2012). The low quality of some institutions of tertiary education has generated an important number of young professionals that can´t find a job. This education segregation, quality of the institutions and lack of public financing were key factors in the 2011 Chilean winter unrest, and to a certain extent has penetrated other domains.

This Chilean case of study is a perfect test bed to study the organizational and temporal dynamics of network generation and diffusion over time and space. We contend that we may both learn and leverage these self-organizing swarms and collaborative networks not only within the educational

## III. CASE STUDY: METHODOLOGY, DATA & ACTORS

The emergence of online social networks opens up unprecedented opportunities to read the collective mind, discovering emergent trends while they are still being hatched by small groups of creative individuals. The Web has become a mirror of the real world, allowing researchers in predictive analytics to study and better understand why some new ideas change our lives, while others never make it from the drawing board of the innovator (Gloor, 2012).

The embedding of the Internet into all aspects of society has led to—among other things—the widespread availability of a cacophony of fact, fiction and opinion composed by all elements of society: governments, news organizations and, most importantly, individuals. Taken together, these players represent a stream of the society's consciousness, albeit a stream with many competing voices, agendas, and noise. From the range of sources, it is clear that the opinions and activities that will move society and make news in the future are embedded in this data stream, tracking in real-time our digital traces (for example using Twitter, Facebook, etc.)

Through a combination of methods, this paper attempts to map and analyze the Chilean student movement in perspective, considering both the macro and the micro network level. To focus the analysis and make it feasible, we have identified a validated set of active actors in this social movement, including 2011-12 student leaders. We also included politicians, government officials, academics and media moguls related to the student movement.

**Tools and Methods**

Through Twitter´s API and public resources such as Twiangulate (see http://www.twiangulate.com/) we have collected relational metadata that uncovers the underlying networks. The relations are constructed through "following" and "followers" and the metadata includes fields like the subjects' self-description, reported real name, number of "followers" and "following", location, etc. Using services such as Klout.com we extracted additional online influence measures and compared the different subjects' style. Since the number of Internet searches related to the student movement can be a powerful and reliable signal of the overall level of interest, we have also used Google Insights to identify trends and potential dependencies. We also use the dynamic social network analysis tool Condor (Gloor & Zhao 2004). Condor collects and analyzes a wide variety of publicly available Web data and provides interfaces supporting various visualizations and interactive data analyses.

Social Media use and Student Movements in Chile
In order to establish a clear pattern between new media platforms and the student "smart mobs" (Reinhold, 2005), we found that in a relatively short period of time there was an important shift in the use of social media. During the 2006 "penguin revolution", the predominant social media platforms used to coordinate action and diffuse information were both blogs and Fotologs. (Garcia, Urbina & Zavala, 2010). Although Internet access was rapidly increasing along with these new social and mobile media tools, yet it was not a spread-out phenomenon. The so-called Web 2.0 (O'Reilly, 2004) platforms were being co-created and gaining momentum in Chile. The contrast with the 2011 context is noticeable, with high Internet access, massive use of Social Networking Sites (SNS) and mobile communication. Five years after, the mobile society (Castells, 2005) was an affordable reality for almost everybody in Chile including social media platforms, which—though evolving—have permeated already everyday life (Silverstone, 2005; Ureta, 2008). This is illustrated in Figure 1, where we show 'interest' and 'search' preferences in social media over time through Google insights.

Figure 1 shows the 2010 tipping point in terms of social media platforms. Twitter is now leaving behind the Fotolog platform, which was massively used for the 2006 Penguin revolution coordination in a country with the higher proportion of Fotolog users in the world (Garcia, Urbina & Zavala, 2010). Facebook became a quite popular SNS and, thus it surpassed Fotolog in mid-2008, showing today almost 30 times more 'interest' rate than Twitter, according to Google Insight.

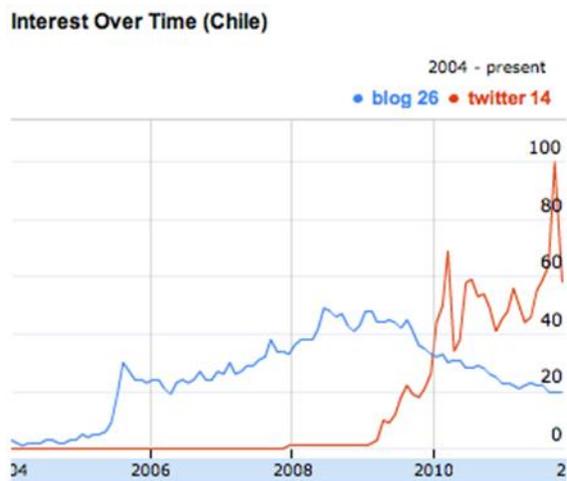

*Figure 1: Popularity of Google Web searches for terms "blog" and "twitter" in Chile*

This is also confirmed by the information provided by Giorgio Jackson and Noam Titelman, presidents of the Catholic University Students' Federation who were key student movement leaders in 2011 and

2012, respectively. They identified Twitter as a major social media platform for the student movement coordination, way ahead of other tools such as blogs and fotolog (appendix B).

This social media platform shift has important consequences. First, communication has become increasingly mobile through high penetration of cell-phones, including affordable smartphones with web connection and direct access to both Facebook and Twitter.

We contend that the horizontal and transparent flow of information under this emerging media system has permeated the culture of the student movement by enacting its self-organizing, and distributed nature. A related interesting discovery is illustrated in Figure 2 where we observe that the relative interest in the word "blog" has fallen drastically in Chile since its peak in June 2008, to only 40% of its original value. As it can be seen in Figure 2, its 'interest rate decrease' coincides with the rise of Twitter, which now has more than double the interest in "blog". This evolving change is, according to Shirky, due to the fact that "The communications tools broadly adopted in the last decade are the first to fit human social networks well, and because they are easily modifiable, they can be made to fit better over time. Rather than limiting our communications to one-to-one and one-to-many tools, which always have been a bad feet to social life, we now have many-to-many tools that support and accelerate cooperation and action" (Shirky, 2009, p.55). In this sense, the Chilean media-enabled social movements evolved from a one-to-many paradigm (Fotolog), to a many-to-many one (Twitter).

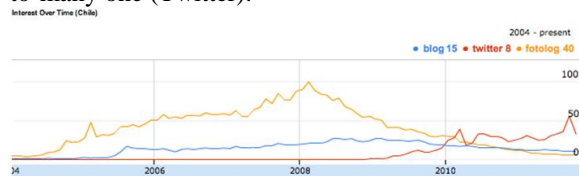

*Figure 2: Popularity of Google Web searches for social media technologies in Chile*

According to our interviews with strategic informants—mainly with two of the leaders of the 2011-2012 Student Movement, see appendix B—, this phenomenon has impacted the leaders of the current movement, who actively use Twitter for both organization and information diffusion. Twitter has been very effective, especially at the moment of promote the marches and other kinds of manifestations. The generalized use of Twitter apps for mobiles allows the smart crowds to be always connected and up-to-date, being a fast-responsive, adaptive network (Rheingold, 2002). The Twitter way offers higher capability for coordination plus an informational advantage when spreading news, but a diminished capability to create elaborated content, hindering the depth of the debate by limiting it to what can fit in 140 characters.

This new use of media emphasizes the immediacy of communication as an evolving stream of information rather than as a cumulative stock of knowledge.

**TimeLine**

To analyze the evolution of the student movement it is necessary to know which have been the key events that have transformed this students' movement into a social movement that has called the attention not only of Chile but of the world. Please see below the key events of the Chilean student movement timeline, which are vital to understand its network topology and evolution through the lens of social media—specially Twitter—use (see appendix D for the complete timeline).

2011
- May 12, 2011: protest called by CONFECH. 20,000 people.
- May 26, 2011: CONFECH deliver a letter with demands of student movement to education minister, Joaquín Lavín.
- June 1, 2011: protest and strike called by CONFECH. 35,000 people.
- June 21, 2011: CONFECH leaders meet with education minister, Joaquín Lavín.
- June 30, 2011: National strike called by CONFECH. 100,000 people.
- July 5, 2011: President Sebastian Piñera and education minister Joaquín Lavín announce in national chain the government proposal: Great National Agreement for the Education (Gran Acuerdo Nacional por la Educación, GANE).
- July 14, 2011: unauthorized march called by CONFECH. 200,000 people across the country.
- July 18, 2011: Felipe Bulnes takes office as education minister.
- August 1, 2011: education minister Felipe Bulnes presents proposal "21 points on education"
- August 4, 2011: violent day of unauthorized mobilizations, calling international attention. More than 800 people arrested in Santiago.
- August 18, 2011: march called by CONFECH . 100,000 people.
- August 24-25: national strike called by Central Union (Central Unitaria de Trabajadores, CUT).

- September 3, 2011: CONFECH, CONES, Teachers Union and CRUCH meets with President Sebastian Piñera and education minister Felipe Bulnes
- September 22, 2011: march called by CONFECH. 100,000 people.
- September 29, 2011, march called by CONFECH. 20,000 people.
- October 12, 2011: CONFECH leaders travel to Europe and meet with international leaders
    - UNESCO
    - Edgar Morin
    - Stéphane Hessel
    - La Sorbonne
    - OCDE
    - ONU
- December 14, 2011: "The Protester", person of the year at TIME
- December 20, 2011: Camila Vallejo, person of the year at The Guardian
- December 29, 2011: Harald Beyer takes office as education minister.

2012
- April 25, 2012: march called by CONFECH. 80,000 people.
- May 16, 2012: march called by CONFECH. 100,000 people.
- August 8, 2012: unauthorized mobilization, called by ACES.
- August 28, 2012: march called by CONFECH. 50,000 people.
- September 24, 2012: CONFECH leaders participate in NUS National Conference, at England.
- September 27, 2012: march called by CONFECH, ACES and CONES. 5,000 people.
- October 17, 2012: CONFECH leaders, representing The Chilean Students Movement, receive The Letelier-Moffitt Human Rights Awards, at Washington DC.

## IV. ANALYSIS: TWITTER DATA & NETWORKS

Our analysis consists of two parts. First, we perform a descriptive analysis of the network topology through six (6) of the most outstanding leaders. Secondly, the main characteristics of the network are analyzed, on the basis of a list of 175 actors who actively partake on the student movement or who are related in an important way to this social movement.

The 6 selected actors for the descriptive analysis are:
- Camila Vallejo: (born in Santiago, April 28, 1988) was a geography student and a leader of the student movement in Chile and the 2011 Chilean protests. She is a prominent member of the Communist Youth of Chile, and was the president of the Student Federation of the University of Chile during 2011 and vice-president of the same institution in 2012.
- Giorgio Jackson: (born in Santiago, February 6, 1987) was a student of Industrial Civil Engineering and leader of the student movement in Chile, including many of the protests. In 2010 he was elected as president of the "FEUC" (Pontifical Catholic University of Chile Student Federation and as spokesman of the "CONFECH" (Chilean Student Confederation).
- Mario Waissbluth: Professor of the University of Chile, President and General Coordinator of "Educación2020". The movement has as objective to create a world-class education system for Chile before 2020.
- Freddy Fuentes: (born in Santiago, October 11, 1993) Spokesman of the "CONES" (High School Students Coordinator).
- José Ancalao: He is werken of the "FEMAE" (Mapuche Student Federation).
- Alfredo Vielma: Spokesman at "ACES" (High School Coordinator Assembly).

Table 1 shows a series of indicators related to the activity in the network of these actors. It's important to notice that these actors (especially the young ones) have passed from practically being unknown to be Chilean-wide celebrities in a few months. Giorgio and Camila have "Twitter similarity" with prominent businesspeople, media celebrities and politicians with long-term careers, such as the current Chilean president, Mr. Sebastián Piñera (measured for example in terms of their number of followers).

We aim to identify relevant people who are influencing the leaders of the student movement. In the "Twittersphere" these are the people who are followed by the actors under study. This motivation comes from observing the ratio of following/followers in Twitter. As is shown in Table 1, the actors have an average ratio of 0,105. This can be interpreted as high selectivity by the key actors who are only following a small number of probably influential agents.

| Actor | Following | Followers | F/F | Tweets | Listed |
|---|---|---|---|---|---|
| Camila Vallejo | 108 | 324730 | 0,0003 | 405 | 1755 |
| Giorgio Jackson | 511 | 137643 | 0,0037 | 2826 | 604 |
| Mario Waissbluth | 581 | 49388 | 0,0118 | 30643 | 785 |
| Freddy Fuentes | 585 | 6752 | 0,0866 | 4478 | 53 |
| José Ancalao | 558 | 5875 | 0,0950 | 2254 | 58 |
| Alfredo Vielma | 86 | 198 | 0,4343 | 90 | 5 |
| Average | 404,8 | 87431,0 | 0,105 | 6782,6 | 543,3 |

*Table 1: Twitter indicators of six key actors*

In the first analysis, we obtained the "top 100" friends of the actors (see appendix A for the selection criteria we used). These results were, however, not satisfactory, mainly because we found mostly famous people such as TV and movie actors, show business people, athletes, etc.

Therefore, we changed the selection criteria of the Twitter friends: instead of selecting the "top 100" we used an algorithm to select an "inner circle of friends". This algorithm is also described in appendix A. It removes the "famous people", giving preference to what might be "closer" friends. We suggest that these people are important to the student movement in two possible ways: their opinion is important to the leaders, or they are part of the movement's core team, acting as knowledge transfer agents or brokers.

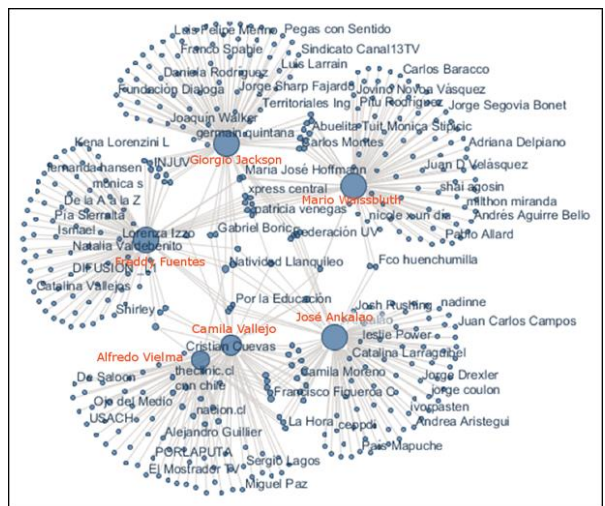

*Figure 3: Social network of six key actors constructed through their Twitter network*

In Figure 3, we see the six main actors selected in this case study: in the periphery we observe their inner circles and ego-networks. In the center of the graph, we observe their shared inner circles forming a star. These actors have two to five of the six main leaders in common.

In order to better understand the online behavior and 'discourse style' of the key leaders of the student movement, we also gathered information from KLOUT, an online tool to "measures influence based on the ability to drive action". This tool uses inputs from Twitter and other social networks to estimate aspects such as reach, amplification and how often top influencers respond to the content that is shared by the selected person.

Using the Klout score we obtained the following information:

| Leader | Score | Klout "style" | True Reach | Amplification | Network |
|---|---|---|---|---|---|
| Camila Vallejo | 75 | Thought Leader | 123K | 7 | 58 |
| Giorgio Jackson | 70 | Pundit | 59K | 7 | 55 |
| Jaime Gajardo | 62 | Broadcaster | 17K | 7 | 47 |
| Mario Waissbluth | 68 | Pundit | 37K | 8 | 53 |
| José Ankalao | 61 | Broadcaster | 14K | 12 | 50 |

*Table 2: Klout influence measures analysis. Explanation of each variable in Appendix C*

Through Table 2, it is possible to shed new light on important dimensions of online presence (mainly twitter) of the key leaders. It becomes clear that Camila Vallejo has a disproportionate amount of "true reach" (and also followers as stated in table 1). Nevertheless, it is important to highlight that this is not directly transferred as amplification (how much the followers retweet or respond to the messages) or the relative influence of the followers (measured through the "network" variable).

The aggregated Klout score also shows that Mario Waissbluth, despite having a significantly smaller "true reach" and smaller number of followers than Vallejo and Giorgio, has a similar aggregated score, due mainly to the amplification of his messages, which are retweeted or responded by his followers. In this case, it could be argued that due to the more seniority and academic nuances of Waissbluth's content, he can amplify his messages across a relatively large and diverse number of influential individuals as well as to generate conversations (instead of simple retweets).

A second part of our analysis was intended to explore the Twitter behavior of a set of 175 actors identified as articulators and leaders of the current student movement (appendix E).

Every actor was classified in 9 categories: Media, Student Leader, Student Movement, Political figure, Social Leader, Government, Institutional Account, and Academics. The results in the sample are the following:

| Tag | Frecuency |
|---|---|
| Academics | 2 |
| Government | 4 |
| Institutional account | 44 |
| Media | 7 |
| Politician | 6 |
| Social Leader | 3 |
| Student Leader | 10 |
| Student Movement | 99 |

*Table 3: Frequency of every category.*

Using this list of actors and the most relevant facts of the timeline, we try to analyze and visualize how the topology and characteristics of the Twitter network relate and eventually anticipate the student movement and demonstrations in the streets. In particular, we analyzed five marches between 2011 and 2012 to compare the network metrics of the central nodes. In addition, we analyze the relationship between the number of tweets in the student movement and the quantity of people who showed up at each mobilization.

We do not rely on a database containing all the tweets from those 175 actors in 2011 and 2012. Though a preliminary analysis of the selected events in both years, this is an attempt to answer the issues mentioned above. The 5 analyzed events are:
- June 30, 2011: National strike called by CONFECH. 100,000 people.
- August 4, 2011: violent day of unauthorized mobilizations, calling international attention. More than 800 people arrested in Santiago.
- September 22, 2011: march called by CONFECH. 100,000 people.
- August 8, 2012: unauthorized mobilization, called by ACES.
- September 27, 2012: march called by CONFECH, ACES and CONES. 5,000 people.

The mobilizations of August 2011 and 2012 stand out for not being authorized marches, which meant high levels of violence, clashes between demonstrators and the police as well as a great deal of violent discourse in Tweeter.

We observe that there are an increasing number of tweets in the network during the day of the mobilization in comparison with the previous week (Table 4). Despite the fact of an incomplete data of 2011, there is a clear trend pointing out to the increase of tweets as the day of the mobilization approaches. Only one of the analyzed events is out of this trend, showing a decrease in the quantity of tweets. Possible explanations for this exception may be related to small quantity of available information for this date or proximity to holidays. Nevertheless, it is possible to state that there is a direct and positive relation between the mobilization in the streets and the activity in the web (Twitter).

| Mobilization date | N° of tweets in the network one week before | N° of tweets in the network in the day of mobilizations | % of increse | Number of assistants to the mobilization |
|---|---|---|---|---|
| 06-30-2011 | 35 | 59 | 69% | 100.000 |
| 08-04-2011 | 14 | 51 | 264% | No data available |
| 09-22-2011 | 25 | 17 | -32% | 60.000 |
| 08-08-2012 | 703 | 957 | 36% | No data available |
| 09-27-2012 | 643 | 689* | 7% | 50.000 |

*Table 4: data from each event*
*\*this number corresponds to the number of tweets in 09-26-2012. No data available for the day of mobilization.*

Besides the change in the number of tweets, we can observe the change in the topology of the network. It is possible to observe a relevant increase in the degree and betweenness centrality.

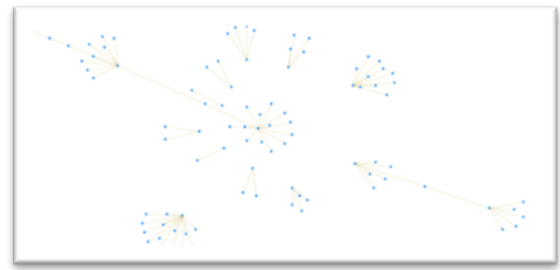

*Figure 4: twitter data before June 30, 2011*

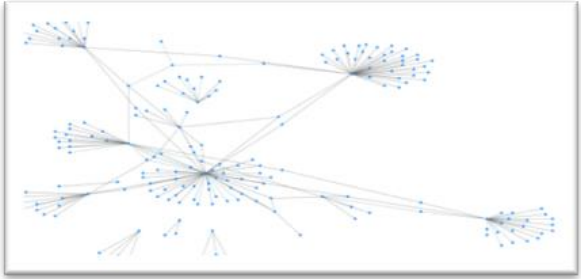

*Figure 5: twitter data after June 30, 2011.*

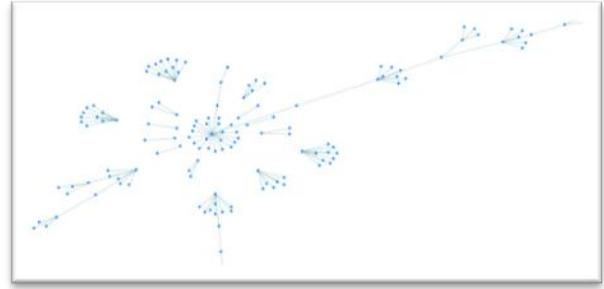

*Figure 9: twitter data after September 22, 2011.*

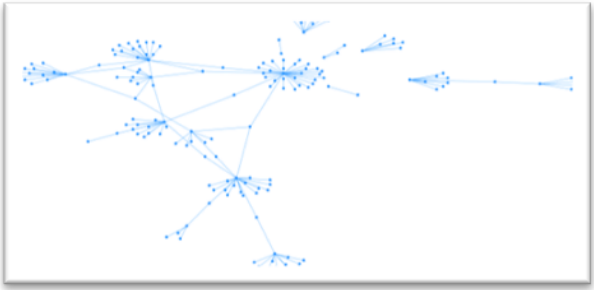

*Figure 6: twitter data before August 4, 2011.*

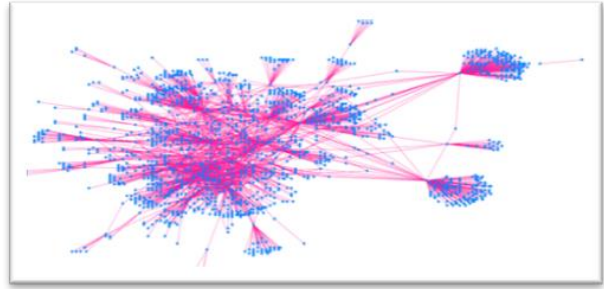

*Figure 10: twitter data before August 8, 2012.*

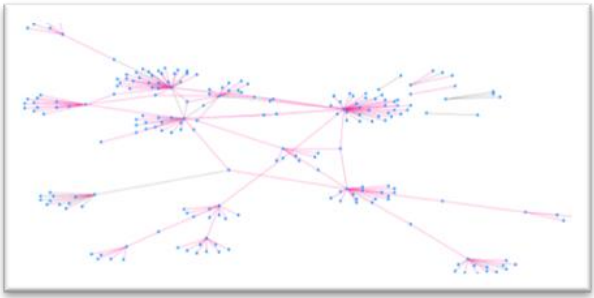

*Figure 7: twitter data after August 4, 2011.*

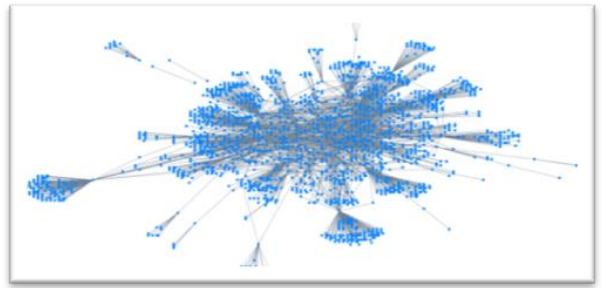

*Figure 11: twitter data after August 8, 2012.*

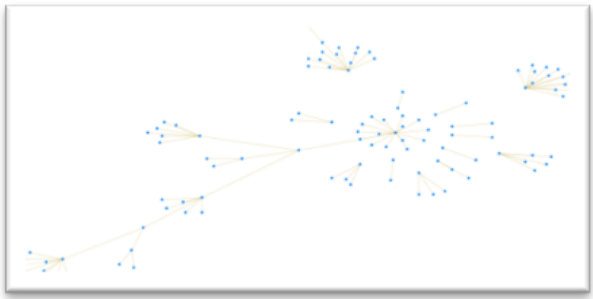

*Figure 8: twitter data after before September 22, 2011.*

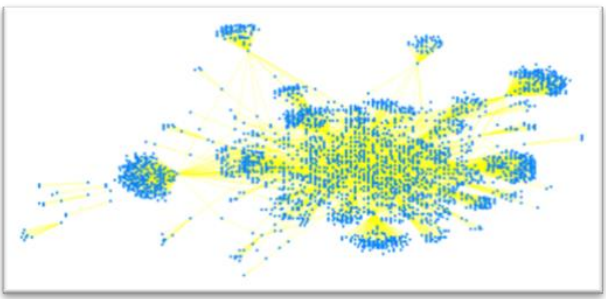

*Figure 12: twitter data before September 27, 2012.*

In addition we analyze the SNA metrics for every event. If we take the top 10 actors in degree centrality, we observe that a few of them appears in more than one event, even from one year to another.

| Degree centrality | | | | |
|---|---|---|---|---|
| 06-30-2011 | 08-04-2011 | 09-22-2011 | 08-08-2012 | 09-27-2012 |
| fireyesr | srcrispin | Fireyesr | sboric | ceciperez1 |
| fernandorojas_ | cesar_reyesp | Srcrispin | obsddhhcl | camilo_b2012 |
| felipesalga | fireyesr | Velagrau | jpluhrs | Biobio |
| srcrispin | pepoglatz | Pepoglatz | difusion_aces | izquierda_tuit |
| m_gallardosoto | mguc | m_gallardosoto | giorgiojackson | Mwaissbluth |
| pepoglatz | felipesalga | Felipesalga | ceciperez1 | el_dinamo |
| cesar_reyesp | velagrau | Vashneo | movilizados2011 | giorgiojackson |
| caroperezd | vashneo | Mineduc | mwaissbluth | cami_carrascoh |
| velagrau | mineduc | Caroperezd | marcoporchile | freddyfuentesm |
| valenzuelalevi | fernandorojas_ | Feuvsantiago | moisesparedesr | movilizatechile |

*Table 5: degree centrality*

| Betweenness centrality | | | | |
|---|---|---|---|---|
| 06-30-2011 | 08-04-2011 | 09-22-2011 | 08-08-2012 | 09-27-2012 |
| fireyesr | srcrispin | Fireyesr | giorgiojackson | ceciperez1 |
| fernandorojas_ | cesar_reyesp | Pepoglatz | sboric | mwaissbluth |
| felipesalga | fireyesr | Mwaissbluth | mwaissbluth | camilo_b2012 |
| tolerancia0 | camila_vallejo | Ankalao | jpluhrs | biobio |
| srcrispin | pepoglatz | Vashneo | difusion_aces | el_dinamo |
| m_gallardosoto | mguc | alexis_gonzlz | ceciperez1 | izquierda_tuit |
| mwaissbluth | c_ballesteros_ | Biobio | gabrielboric | giorgiojackson |
| pablochapav | felipesalga | Fepucv | obsddhhcl | cami_carrascoh |
| pepoglatz | vashneo | Piaeberhard | marcoporchile | sebagamboa_ freddyfuentesm |
| paul_guzman | panchopinochet | Mineduc | movilizados2011 | |

*Table 6: betweenness centrality*

The fact that the actors with major links inside the network repeat themselves across the different events shows that there are certain leaders who influence the online (Tweeter) space of this student movement in a permanent way. Some of them are student leaders, some academic and political actors. This fact confirms this movement is not only a student movement but increasingly a social networked one. Actually, we observe an increase of the politicians who are located in central and influential domains of this network.

The actors who appear with high betweenness centrality are, in most cases, those who appear also with high degree centrality. That is to say, those who generate more connections among different nodes (bridging function) are those who are more directly connected with others (bonding function).

### V. PRELIMINARY CONCLUSIONS

The 2011-2012 Chilean Student Movement offers an excellent test bed to study the dynamics of network generation and diffusion through social media, specially Twitter. We encounter these self-organizing swarms and collaborative networks not only within the educational space, but also in the broader political space. That is to say, we have witnessed the path from a student movement towards a broader social movement, and –by the time of writing—to new political associations lead by Giorgio Jackson that may become political parties in the near future.

Our analysis of metadata obtained from Twitter profile information illustrates the relevance, interconnectivity and entanglement among the movement's leaders, political representatives, the media and key institutional agents. This collective student-based movement has, in fact, infected both the political and media systems by mobilizing resources, information and people across the public sphere, combining the digital and physical world.

Despite the fact of the lack of some information for 2011, we were able to point out to the direct and positive relationship between the activity in the Tweeter network and the activity in the streets, as observed in the 5 selected events in 2011 and 2012.

Finally, we contend that Twitter is an excellent source to track the dynamics of both organization, diffusion and evolution of the student movement as described by the leaders and verified in our digital data set, including 175 actors for selected events in 2011 and 2012.

**Further work**

In the near future, we expect to conduct interviews to other leaders of the movement, with in-depth questions aiming to understand the meaning of this social media-enabled student movement as well as to understand the media repertoire to coordinate and foster collective action. In addition, we will continue tracking the social media metrics and temporal network evolution of this movement as it unfolds over time, including leadership change and entanglement with the broader political space.

Finally, we will include semantic analysis of both Tweeter and Wikipedia to observe the evolution and change of the positions, demands and counter-offers in the negotiations between the student movement, the government, Universities, Parliament and political parties. We will also perform a sentiment analysis of its Tweeter network, to get further information about the content and discursive nature of this social movement in Chile, including the political and higher education agents as well as to anticipate its development over time.

## VI. APPENDIX

**Appendix A: Definitions**
- Top 100 selections: This tool selects a maximum of 100 key followers of the actor. Key followers are calculated based on the influence and their own followers.
- Inner Network: Usually people with less followers but closer to the actor. These are people who might go undetected because they don't have a large number of followers on Twitter.

**Appendix B: Giorgio Jackson's and Noam Titelman's answers:**

*Giorgio Jackson*

Using a scale from 1 to 7, where 1 is the lowest evaluation and 7 the highest; assign a score to the media with more impact in the **organization** (strategy, logistics, fundraising, etc.) of the student movement, which you have led:

| Media | Score |
|---|---|
| Facebook | 5 |
| Twitter | 6 |
| Email | 5 |
| Mobile | 4 |
| Blogs | 2 |
| Traditional media(newspapers, television, radio, etc) | 3 |

Using a scale from 1 to 7, where 1 is the lowest impact and 7 the highest; assign a score to the media with more impact in the **diffusion** (strategy, logistics, fundraising, etc.) of the student movement.

| Media | Score |
|---|---|
| Facebook | 6 |
| Twitter | 6 |
| Email | 4 |
| Mobile | 3 |
| Blogs | 4 |
| Traditional Media(newspapers, television, radio, etc) | 7 |

*Noam Titelman*

Using a scale from 1 to 7, where 1 is the lowest evaluation and 7 the highest; assign a score to the media with more impact in the **organization** (strategy, logistics, fundraising, etc.) of the student movement which you have led:

| Media | Score |
|---|---|
| Facebook | 3 |
| Twitter | 3 |
| Email | 6 |
| Mobile | 5 |
| Blogs | 1 |
| Traditional Media(newspapers, television, radio, etc) | 2 |

Using a scale from 1 to 7, where 1 is the lowest evaluation and 7 the highest; assign a score to the media with more impact in the **diffusion** (strategy, logistics, fundraising, etc.) of the student movement which you have led:

| Media | Score |
|---|---|
| Facebook | 4 |
| **Twitter** | 5 |
| **Email** | 3 |
| **Mobile** | 3 |
| **Blogs** | 1 |
| **Traditional Media(newspapers, television, radio, etc)** | 7 |

## Appendix C: Klout variables
- **True Reach:** How many people you influence
- **Amplification:** How much you influence them
- **Network Impact:** The influence of your network

## Appendix D: Time Line
**2011**
- o May 12, 2011: protest called by CONFECH. 20,000 people.
- o May 26, 2011: CONFECH deliver a letter with demands of student movement to education minister, Joaquín Lavín.
- o May 30, 2011: CONFECH leaders meet with education minister, Joaquín Lavín.
- o June 1, 2011: protest and strike called by CONFECH. 35,000 people.
- o June 16, 2011: unauthorized protest called by CONES.
- o June 21, 2011: CONFECH leaders meet with education minister, Joaquín Lavín.
- o June 30, 2011: National strike called by CONFECH. 100,000 people.
- o July 5, 2011President Sebastian Piñera and education minister Joaquín Lavín announce in national chain the government proposal: Great National Agreement for the Education (Gran Acuerdo Nacional por la Educación, GANE).
- o July 14, 2011: unauthorized march calleb by CONFECH. 200,000 people across the country.
- o July 18, 2011: Felipe Bulnes takes office as education minister.
- o August 1, 2011: education minister Felipe Bulnes presents proposal "21 points on education"
- o August 4, 2011: violent day of unauthorized mobilizations, calling international attention. Mora than 800 people arrested in Santiago.
- o August 18, 2011: march called by CONFECH ("Marcha de los paraguas"). 100,000 people.
- o August 24-25: national strike called by CUT
- o September 3, 2011: CONFECH, CONES, Teachers Union and CRUCH meets with President Sebastian Piñera and Education Minister Felipe Bulnes
- o September 22, 2011: march called by CONFECH. 100,000 people.
- o September 29, 2011, march called by CONFECH. 20,000 people.
- o October 12, 2011: CONFECH leaders travel to Europe and meet with international leaders
    - UNESCO
    - Edgar Morin
    - Stéphane Hessel
    - La Sorbonne
    - OCDE
    - ONU
- o October 15, 2011: global mobilizations 15-O. Indignados, Occupy Wall Street
- o October 18-19, 2011: national mobilization. 200,000 people across the country.
- o December 14, 2011: "The Protester" person of the year at TIME
- o December 20, 2011: Camila Vallejo, person of the year at The Guardian
- o December 29, 2011: Harald Beyer takes office as education minister.

**2012**
- o April 25, 2012: march called by CONFECH. 80,000 people.
- o May 16, 2012: march called by CONFECH. 100,000 people.
- o August 8, 2012: unauthorized mobilization, called by ACES.
- o August 28, 2012: march called by CONFECH. 50,000 people.
- o September 6, 2012: CONFECH leaders meet education minister Harald Beyer
- o September 24, 2012: CONFECH leaders participate in NUS National Conference, at UK.
- o September 27, 2012: march called by CONFECH, ACES and CONES. 5,000 people.
- o October 17, 2012: CONFECH leaders, representing The Chilean Students Movement, receive The Letelier-Moffitt Human Rights Awards, at Washington DC.

**Appendix E: 175 actors list**

| Name | Handle |
|---|---|
| #Movilizate | @Movilizatecl |
| Acción UNAB! | @AccionUNAB |
| ACES | @Difusion_ACES |
| Aintzane Lorca | @AintzaneLorca |
| Alberto Mayol | @AlbertoMayol |
| Alexis J. Gonzalez | @alexis_gonzlz |
| Alfredo Vielma | @alfredovielmav |
| Andrés Fielbaum | @Afielbaum |
| Andrés Velasco | @AndresVelasco |
| Angello Giorgio | @AngelloGiorgio |
| Bárbara Figueroa | @Barbara_figue |
| Bárbara Vallejos | @LalaVallejos |
| Cadena Nacional | @cadena_nacional |
| Camila Carrasco H. | @cami_carrascoh |
| Camila Carvallo | @camicarvallo |
| Camila Donato | @DonatoCamila |
| Camila Vallejo | @camila_vallejo |
| Camilo Ballesteros | @camilo_B2012 |
| Camilo Riffo | @camiloriffo |
| Carla Amtmann | @Carla_Amtmann |
| Carlos Alarcón | @carlconm |
| Carlos Figueroa | @CarlosFigue |
| Carmela_movilizada | @ccp_movilizada |
| Carolina Jara | @Carolina_Jarap |
| Carolina Pérez | @caroperezd |
| Catalina Lamatta | @cata_lamatta |
| Cecilia Pérez Jara | @ceciperez1 |
| César Reyes | @cesar_reyesp |
| Claudio Orrego | @orrego |
| Colegio de Profesores | @MagisterioNac |
| Comunicaciones FECh | @la_fech |
| Comunicacionesfeuach | @feuach_comunica |
| CONES | @CoNESChile |
| CONFECH | @confech |
| Cooperativa.cl | @cooperativa |
| Crecer UC | @creceruc |
| Cristián Andrade | @andradecristian |
| Cristián Stewart | @cstewartc |
| Cristóbal Lagos | @cristobalturron |
| Cristofer Sarabia | @Cris_Sarabia |
| Danae Díaz Jeria | @DanaeDiazJeria |
| Daniela López | @DanielaLopezLv |
| Daniela Serrano | @SerranoDaniela_ |
| David Terzán | @Dterzan |
| defensoriapopular | @defenspopular |
| Diego Schalper | @dischalper |
| Diego Vela | @velagrau |
| Educación 2020 | @Educacion2020 |
| El Dínamo | @el_dinamo |
| El Mostrador | @elmostrador |
| El Pingüino Informa | @Pinguinoinforma |
| El Post | @elpost_cl |
| Elecciones FECH | @Elecciones_FECH |
| Elecciones FEUC | @EleccionesFeuc |
| Elías Lonconado | @EliasLonconado |
| Eloísa Gonzáles | @_EloisaGonzalez |
| Emilia Malig | @emiliamalig |
| Estafados CORFO | @Estafados_CORFO |
| Estudiante informado | @infestudiantes |
| FECH | @la_fech |
| Fel Stgo | @FEL_Stgo |
| Felipe Mery Cardoza | @FelipeMeryC |
| Felipe Ramírez | @feliperasa |
| Felipe Salgado | @Felipesalga |
| Felipe Valdebenito | @pipovaldebenito |
| FEMAE | @FedFEMAE |
| FEPUCV | @FEPUCV |
| Fernanda Sandoval | @FdaSandoval |
| Fernando Reyes | @fireyesr |
| Fernando Rojas | @fernandorojas_ |
| FEUA | @FEUAntofagasta |
| FEUACH | @FeuachUACH |
| FEUAI 2012 | @FEUAI_stgo |
| FEUANDES | @FEUANDES |
| FEUBB | @feubb |
| FEUBO | @FEUBO_OFICIAL |
| FEUC | @feuc |
| FEUCEN | @_FEUCEN |
| FEUCM 2011 | @FEUCM2011 |
| FEUCN | @FEUCN |
| FEUCN-Coquimbo | @feucncqbo |
| FEUDD | @FEUDD_stgo |
| FEULS | @Feuls |
| FEUPLA | @Feupla |
| FEUSACH | @feusach |
| FEUSAM | @FEUSAM |
| FEUSMJMC | @feusmjmc |
| FEUST Santiago | @FeustSantiago |
| FEUTEM2012 | @FEUTEM |
| FEUTFSM | @feutfsm |
| FEUV | @feuv |
| FEUV Santiago | @feuvsantiago |
| FEUVM | @cegesuvm2012 |
| Francisco Figueroa | @panchofigueroa |
| Francisco Fuenzalida | @ffuenza |
| Francisco Pinochet | @Panchopinochet |
| Franco Parisi | @Fr_parisi |
| Freddy Fuentes Matth | @FreddyFuentesM |
| Gabriel Boric | @gabrielboric |
| Gabriel González C | @GabGonzalezC |
| GaryPasténHermosilla | @Gary_mbb_ubo |
| Gastón Urrutia | @GastonUrrutiaC |
| Giorgio Jackson | @giorgiojackson |
| Giovanna Roa | @giovannaroa |
| Gustavo Pacheco | @TavoPacheco_ |
| Ignacio Saffirio | @isaffirio |
| Izquiera Tuitera | @Izquierda_Tuit |
| Izquierda Autónoma | @izqautonoma |
| Jaime Gajardo | @jaimegajardo |
| Javier Miranda | @j_miranda_s |
| JJCC | @jjcc_chile |
| Joaquín Walker | @JoaquinWalker |
| Jorge Brito | @jorbritoh |

| Name | Handle |
|---|---|
| José Ankalao | @ANKALAO |
| José Antonio Gómez | @jagomez |
| José Manuel Morales | @jjosemanuel |
| José Vidal | @jm_vidal |
| Juan Pablo Luhrs | @jpluhrs |
| Julio Maturana | @julio_maturana |
| Julio Sarmiento | @sarmiento510 |
| Karol Cariola | @Karolcariola |
| Luna Violeta | @LunaMorales |
| MACE UAI | @MACEUAI |
| Manuel Gallardo | @M_GallardoSoto |
| Manuel Palma | @ManuelPalmaM |
| MarcoEnriquezOminami | @marcoporchile |
| María del Pilar Gras | @pilargras |
| Mario Waissbluth | @mwaissbluth |
| Marjorie Cuello | @marjoriecuello |
| Marta Lagos | @mmlagoscc |
| Matías Reeves | @MatiasReeves |
| MG | @mguc |
| Miguel Crispi | @srcrispin |
| MINEDUC | @Mineduc |
| Moisés Paredes | @MoisesParedesR |
| Movilizados 2011 | @Movilizados2011 |
| MovilizateChile | @MovilizateChile |
| Nataly Espinoza | @NatalyEspin0za |
| Nicolás Gajardo | @NicoGajardo_ |
| Nicolás Valenzuela | @valenzuelalevi |
| Noam Titelman | @NoamTitelman |
| Nueva Acción Universitaria (NAU) | @naupuc |
| Observadores DDHH CL | @obsDDHHcl |
| Pablo Chamorro | @pablo_fepucv |
| Patricio Indo | @Patricio_Indo |
| Patricio Órdenes | @PatricioOR |
| Paul Floor Pilquil | @vashneo |
| Pedro Glatz | @pepoglatz |
| Privadas | @privmovilizadas |
| Quenne Aitken | @QuenneAitken |
| Rdemocrática | @Rdemocratica |
| Recaredo Gálvez | @recarex |
| RED | @red_puc |
| Red SurDA | @movimientosurda |
| René Andrade | @reneancar |
| Rodrigo Echecopar | @raecheco |
| Rodrigo Hinzpeter | @rhinzpeter |
| Rodrigo Rivera | @Rodrigo_RiveraC |
| Rodrigo SalazarJ | @Rodrigofsj |
| Scarlett Mac-Ginty | @ScarlettMac |
| Sebastián Donoso | @Sdonoso_ |
| Sebastián Farfan S. | @sebafarfans |
| Sebastián Gamboa | @sebagamboa_ |
| Sebastián Godoy | @SebaGodoyElguet |
| Sebastián Vielmas | @sebavielmas |
| Simon Ballesteros | @simon_ballest |
| Simón Boric | @sboric |
| Solidaridad UC | @SolidaridadUC |
| UNE Chile | @une_chile |
| Valentina Latorre | @vlatorre |
| Ximena Rincón | @ximerincon |
| Yoxcy Campos | @YoxcyCampos |
| Bio Bio Chile | @biobio |
| La Tercera | @latercera |
| El Mercurio | @emol |

## ACKNOWLEDGEMENTS

The authors are grateful to Francisco Moya, Javier Urbina, and Patricia Hansen for their research input and suggestions. We also want to thank Giorgio Jackson and Noam Titelman, student leader, for their invaluable help and disposition to answer our questions.